\documentclass[aps,pre,reprint,groupedaddress,showpacs]{revtex4-1}

\usepackage{graphicx} 
\usepackage{comment}
\usepackage{color}

\begin{document}


\title{Reduced mobility of infected agents suppresses but lengthens disease in biased random walk}


\author{Genki Ichinose}
\email[]{ichinose.genki@shizuoka.ac.jp}
\altaffiliation{}
\affiliation{Department of Mathematical and Systems Engineering, Shizuoka University, Hamamatsu 432-8561, Japan}

\author{Yoshiki Satotani}
\affiliation{Department of Information Technology, Okayama University, Okayama 700-8530 Japan}

\author{Hiroki Sayama}
\altaffiliation{}
\affiliation{Center for Collective Dynamics of Complex Systems \\Department of Systems Science and Industrial Engineering, Binghamton University, State University of New York, Binghamton, NY 13902-6000}

\affiliation{School of Commerce, Waseda University, Tokyo 169-8050, Japan}

\author{Takashi Nagatani}
\affiliation{Department of Mechanical Engineering, Shizuoka University, Hamamatsu 432-8561, Japan}


\date{\today}

\begin{abstract}
Various theoretical models have been proposed to understand the basic nature of epidemics.
Recent studies focus on the effects of mobility to epidemic process.
However, uncorrelated random walk is typically assumed as the type of movement.
In our daily life, the movement of people sometimes tends to be limited to a certain direction, which can be described by biased random walk.
Here, we developed an agent-based model of susceptible-infected-recovered (SIR) epidemic process in a 2D continuous space where agents tend to move in a certain direction in addition to random movement.
Moreover, we mainly focus on the effect of the reduced mobility of infected agents.
Our model assumes that, when people are infected, their movement activity is greatly reduced because they are physically weakened by the disease.
By conducting extensive simulations, we found that when the movement of infected people is limited, the final epidemic size becomes small.
However, that crucially depended on the movement type of agents.
Furthermore, the reduced mobility of infected agents lengthened the duration of the epidemic because the infection progressed slowly.
\end{abstract}


\maketitle


 \section{Introduction}
Epidemics refer to the state in which an infectious disease spreads extensively and rapidly to a large number of individuals.
Once a pandemic happens, many people seriously suffer.
To understand the basic nature of epidemics, various mathematical models have been developed.
When an infection occurs, contacts between susceptible and infectious individuals are required.
Traditional mathematical epidemic models assume that the contact process takes place without explicit spatial structure where homogeneous mixing of susceptible and infectious individuals is considered \cite{KermackMcKendrick_ProcRSocA_1927, Hethcote_SIAMRev_2000, Diekmann_etal2012, Allen_etal2008}.

In contrast with those traditional models, the spatio-temporal distribution of susceptible and infectious individuals and their pattern of contacts are important to analyze and predict epidemic spreading of populations with spatial structures in humans.
In this case, contacts between susceptible and infectious individuals are caused by the movements of individuals.
For this reason, epidemic models with explicit individual movement have attracted much attention \cite{Steven_etal_PLoSNeglTropDis2009, Barmak_etal_PhysRevE2011, Li_etal_EPJB2010, Zhou_etal_PhysRevE2012, Hong_etal_IntJSystSci2016, Buscarino_etal_PhysRevE2014, Gong_etal_PhysA2014, Barmak_etal_PhysA2016, Li_etal_CommunNonlinearSci2015, Buscarino_etal_IntJBifurcationChaos2010, Nagatani_etal_JTheorBiol2018, Huang_etal_JStatMechl2016, SharmaGupta_PhysAl2017}.
First, epidemic models with uncorrelated random walk have been considered \cite{HosonoIlyas_MathModelsMethods1995, Wang_etal_DisConDynSys2012, Li_etal_CommPureApplAnal2015, BaiZhang_CommNonlinSciNumerSimul2015}.
Those models were often described by partial differential equations (PDE).
There are many situations that the movement of people is not described by uncorrelated random walk.
For instance, if there are some obstacles in their living environments, people have to avoid those obstacles when they move.
People tend to move on a street rather than randomly moving across the ground.
Moreover, when people walk to commute, they tend to head toward a specific destination such as a station.
This directional walking pattern can be described by a biased random walk.
There are some PDE-based epidemic models which deal with the biased random walk \cite{BeardmoreBeardmore_ProcRSocLondA2003, GudeljWhite_TheorPopulBiol2004, Gudelj_etal_BullMathBiol2004}.

On the contrary, agent-based models (ABM) with explicit movement have also been proposed to describe epidemic spreading \cite{ZhouLiu_PhysA2009, Pan_JSTAT2016, Zhou_PhysA2009, Buscarino_EPL2008, Nagatani_etal_JPSJ2017, Buscarino_etal_PhysRevE2014}. 
Compared to PDE models, ABM has an advantage in the sense that detailed dynamics of epidemic spreading can be easily described, simulated and tracked.
For example, by using ABM, we can know when and where infection occurs, which is important for preventing disease from spreading.
Therefore, we use it for the modeling of epidemic spreading.
In those ABMs, directional movement of agents is not considered except for Ref.~\cite{Nagatani_etal_JPSJ2017}.
Nagatani et al.~\cite{Nagatani_etal_JPSJ2017} studied an SIS (susceptible-infected-susceptible) model combined with  the directional movement of agents in a one-dimensional space.
They showed that there is a critical density whether epidemics spread or not. If the density is larger than the critical density, epidemics never disappear because contacts between infectious and susceptible individuals take place in a crowd like a traffic jam.
In contrast, if the density is lower than the critical limit, epidemics finally disappear.
However, the model is relatively simple because the space is a one dimensional line segment.

Here, we extend the model to a two-dimensional space.
More importantly, we consider the effect of the reduced mobility of infected agents.
When people are infected, their movement activities are greatly reduced because they are physically weakened by the disease.
In a one-dimensional space, this effect promotes disease spreading because it causes a traffic jam.
It is not obvious when the dimension is extended.
In this paper, we focus on the effect of reduced mobility to disease spreading in a 2D continuous space where agents adopt a biased random walk.
 
 \section{Model}
 \label{sec:Model}
We consider the situation that $N$ agents interact with each other by moving on a 2-D continuous space with periodic boundary conditions.
The size of the space is $L \times L$.
Each agent does not have a size and is considered a particle.
The infection and recovery processes take place based on the SIR model.
Thus, the state of each agent can be either susceptible ($S$), infected ($I$), or recovered ($R$), which is specified by $\eta \in \{S,I,R\}$.

Each agent is selected in random order and the following two phases are conducted for each agent. 
First, the movement phase takes place. To reveal the effect of biased movements, agents move to a certain direction (rightward) not only random diffusion.
Moreover, the movement of $I$ is different from the one of $S$ and $R$ because we assume $I$'s activity decreases due to illness.
Specifically, the location of agent $i$ at time $t+1$ in the state $\eta$ is given by
\begin{eqnarray}
x_{i,\eta}(t+1)&=&x_{i,\eta}(t)+N(0, \sigma_\eta^2)+\epsilon_\eta, \nonumber \\
y_{i,\eta}(t+1)&=&y_{i,\eta}(t)+N(0,\sigma_\eta^2),
\end{eqnarray}
where $x$ and $y$ specify the coordinates of the agent on the continuous space.
$N(0, \sigma_\eta^2)$ is the normal distribution with mean 0 and standard deviation $\sigma_\eta$.
$\epsilon_\eta$ denotes the advection velocity which leads to biased movement.
We set $\sigma_\eta=1$ and $\epsilon_\eta=1$ when $\eta=S$ or $\eta=R$.
Conversely, because the movement of $I$ is limited due to illness, we set $0 \leq \sigma_I \leq 1$ and $0 \leq \epsilon_I \leq 1$ for $\eta=I$.

Second, the epidemic phase takes place. $S$ becomes $I$ when the Euclidean distance between them is less than or equal to the infection radius $r$.
This means that infection always takes place within the radius. In other words, infection rate is 1. 
$I$ recovers from infection and becomes $R$ with the recovery rate $\gamma$.
Once recovered, $R$ never gets infected.
Thus, the fraction of $R$ after the disease completely disappears can be considered as the final epidemic size, which we will mainly focus on in the results.

In the simulations, we define one Monte Carlo step when every agent has been selected once.
Thus, the iteration proceeds asynchronously.
We vary $\sigma_I$ and $\epsilon_I$ as the main experimental parameters and study how the final epidemic size and the extinction time of disease change depending on those parameters.

 \section{Results and Discussions}
 \label{sec:Result}
\begin{figure*}[!t]
  \centering
  \includegraphics[width=\textwidth]{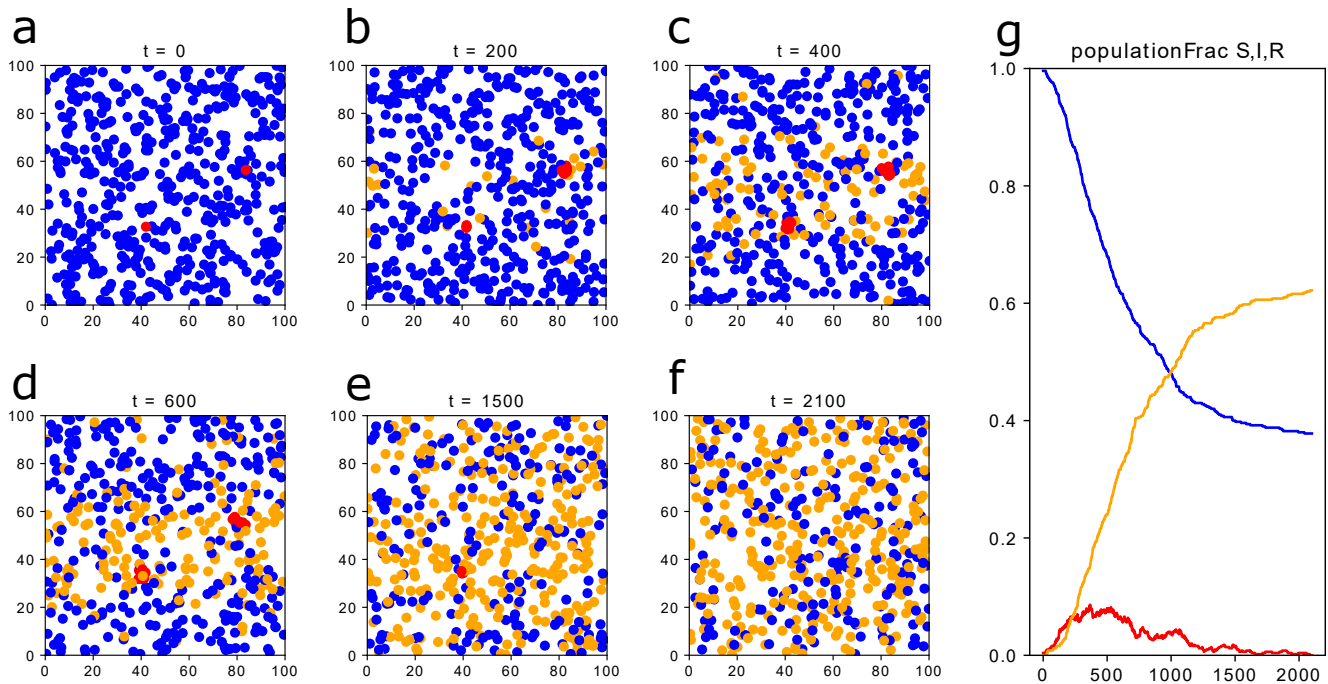}
  \caption{(Color online) The result of the no movement case ($\sigma_I=0, \epsilon_I=0$). $\gamma=0.01$.
   (a)-(f) Snapshots of the epidemic spreading. (g) Population dynamics of the epidemic spreading. Blue, red, and orange denote susceptible, infectious, and recovered agents, respectively.}
  \label{noMovement}
 \end{figure*}

 We used the following parameter setting unless noted otherwise: $L=100, N=500$, and $r=2$.
 In $N$ agents, the number of the initial infected agents is $I_0 =2$ (We also checked the cases of $I_0=1$ and 3. See the supplemental materials).
 Initially, $N$ agents are randomly distributed in the 2-D space  (See Fig.~\ref{noMovement}a and the supplemental movie).

\subsection{Two limiting cases}
\begin{figure*}[!t]
  \centering
  \includegraphics[width=\textwidth]{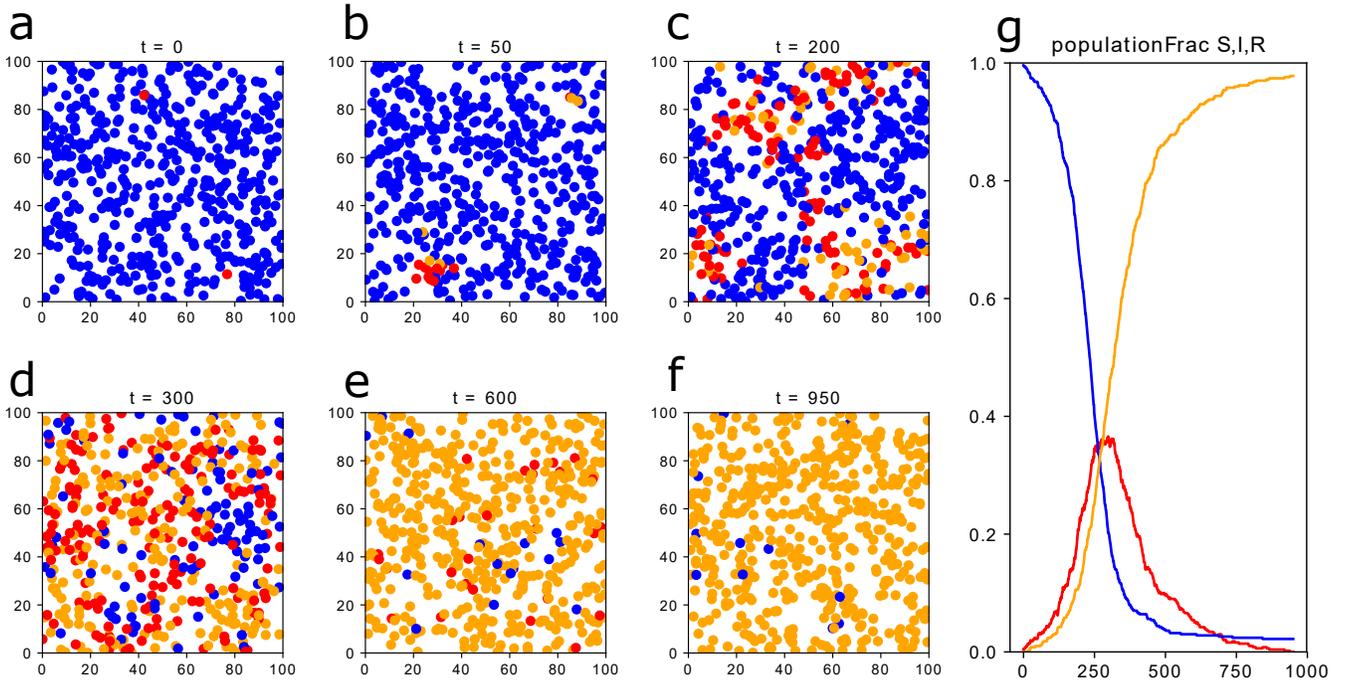}
  \caption{(Color online) The result of the high movement case  ($\sigma_I=1, \epsilon_I=1$). In this setting, there is no difference in the movement of infected, susceptible, and recovered agents. The other settings are the same as Fig.~\ref{noMovement}.}
  \label{highMovement}
 \end{figure*}

First, we consider the two limiting cases, which are $\sigma_I=0, \epsilon_I=0$ and $\sigma_I=1, \epsilon_I=1$.
The former corresponds to our main focus where infected agents do not move at all. We call this the \textit{no movement case} hereafter.
This is relatively unrealistic. Therefore, we will relax this condition where the movement of infected agents is small but non-zero. 
On the contrary, in the latter, there is no difference among $S, R$ and $I$ about the movement. We call this the \textit{high movement case} hereafter.
However, there are great differences between the latter case and the traditional epidemic mean-field model because we assume spatial structure and directional movements in our model.
Figures \ref{noMovement} and \ref{highMovement} show the corresponding simulation results in one run, respectively.
See also the movies for each simulation in the supplemental materials.

In the no movement case (Fig.~\ref{noMovement}), infection only occurs when susceptible agents pass through infectious agents.
When susceptible agents get infected, they stop there.
Also, there seems to be stripes around the infected agents until Fig.~\ref{noMovement}d.
However, these stripes collapse as time goes by because recovered agents move around the space (Fig.~\ref{noMovement}e), and finally disease disappears by $t=2100$ (Fig.~\ref{noMovement}f).
The final epidemic size (the final fraction of recovered agents) is about 0.63 (final point of orange line in Fig.~\ref{noMovement}g).
The epidemics do not spread throughout the space because the location of infection is limited.

Figure \ref{highMovement} shows the result of the high movement case.
In this case, there is no difference in the movement of infected, susceptible, and recovered agents.
Thus, the epidemic immediately spreads throughout the space due to the movement of infected agents (See Fig.~\ref{highMovement}c and d).
Once that happens, the disease gradually decreases and finally disappears (Fig.~\ref{highMovement}f).
The final epidemic size is about 0.98, which means almost all agents got infected (Fig.~\ref{highMovement}g).
This is because contacts between infectious and susceptible agents frequently occur because of the high movement of infected agents.
Therefore, the final epidemic size is much larger than the no movement case (Fig.~\ref{noMovement}).
Another interesting difference between them is that, compared to the no movement case, the extinction time of the disease is less than half in the high movement case.
If we compare Figs.~\ref{noMovement}g and \ref{highMovement}g, we see that the infection proceeds slowly in the no movement case.
In this case, the location of infection is quite limited, resulting in the existence of the disease for a long time.
We will see the detailed effect of inactivity and directionality of infected agents in the next section by conducting a sensitivity analysis.

\subsection{Sensitivity analysis}
\begin{figure}[!t]
  \centering
  \includegraphics[width=\columnwidth]{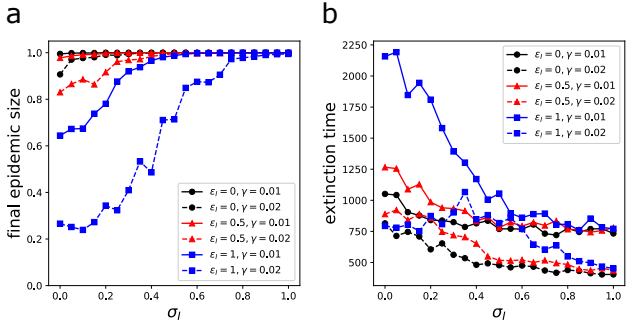}
  \caption{(Color online) The effect of  $\sigma_I, \epsilon_I$, and $\gamma$ for the final epidemic size and the extinction time of disease.}
  \label{sensitivityAnalysis}
 \end{figure}

To reveal the effect of the two main parameters, $\sigma_I$ and $\epsilon_I$ in detail, we conducted a sensitivity analysis for them.
Figure \ref{sensitivityAnalysis} shows the final epidemic sizes and the extinction time of disease when $\sigma_I, \epsilon_I$, and $\gamma$ are changed.
In these results, each simulation is run until the disease completely disappears, and  two hundred simulation runs are averaged for each data point.
Basically, when $\gamma$ is high, the final epidemic size is suppressed at lower values.
This is simply because, when $\gamma$ is high, infected agents tend to recover before the further infection occurs.

Next, we focus on the effect of $\sigma_I$. As $\sigma_I$ becomes larger, the final epidemic size increases.
This is consistent with the results in Figs.~\ref{noMovement} and \ref{highMovement}.
$\sigma_I$ means the diffusion rate of infected agents. If the rate is high, the infected agents move around the entire space.
Therefore, it raises the final epidemic size.

\begin{figure*}[!t]
  \centering
  \includegraphics[width=\textwidth]{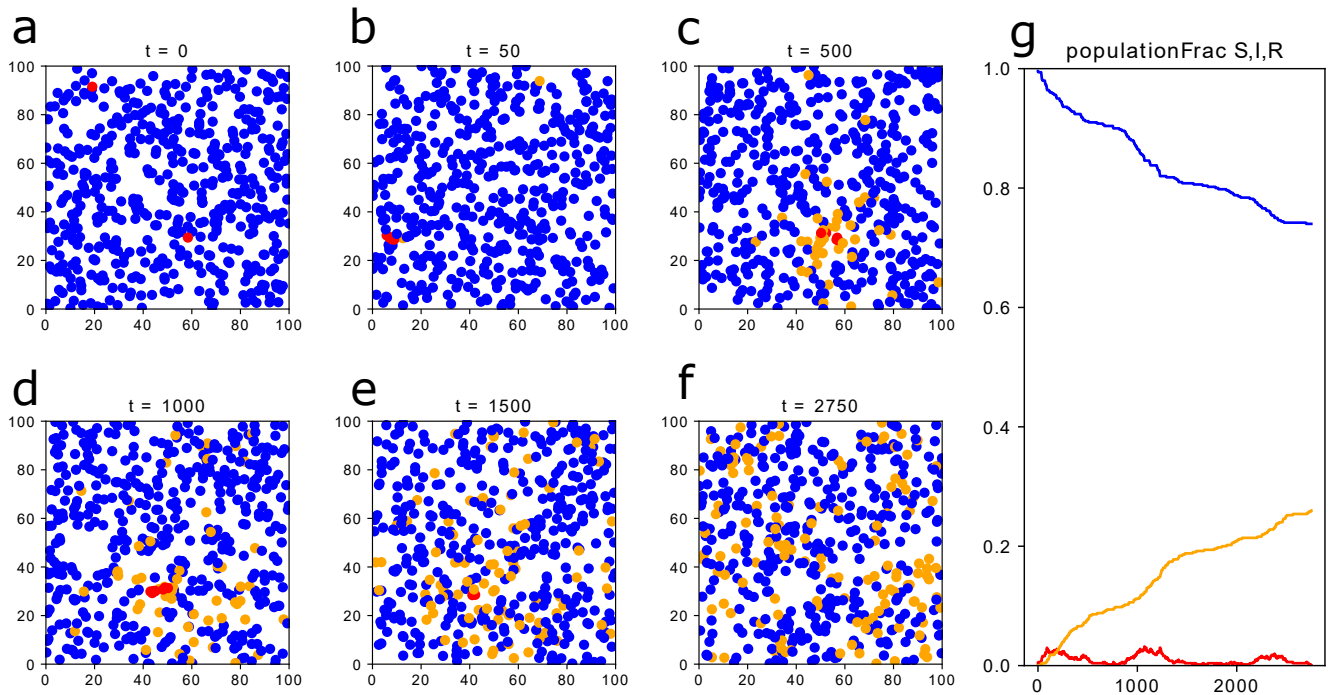}
  \caption{(Color online) The result of the minimum contact case ($\sigma_I=0, \epsilon_I=1$). In this setting, the contact opportunities with the infected agents are very rare, which leads to the lowest final epidemic size. The other settings are the same as Fig.~\ref{noMovement}.}
  \label{biasMovement}
 \end{figure*}

In contrast, the effect of $\epsilon_I$ is different than $\sigma_I$. The difference between $\epsilon_I=0$ and $\epsilon_I=1$ when $\sigma_I=0$ is significant.
Thus, we compare the results between them in detail.
Figure \ref{biasMovement} shows the screenshots and population dynamics at $\sigma_I=0$ and $\epsilon_I=1$.
In this setting, although infected agents do not perform random diffusion, they move rightward with the other types of agents ($S$ and $R$) at the same speed, which means the relative advection velocity is 0.
Therefore, only a few agents around the infected agents have a chance to get infected.
In other words, the contact opportunities with the infected agents are very rare. We call this the \textit{minimum contact case}.
In the minimum contact case, the advection velocity of $I$'s agent is the same as the one in $S$'s and $R$'s agents.
This leads to the lowest final epidemic size.
In contrast, in the no movement case (Fig.~\ref{noMovement}), there is still a higher chance of contact with the infected agents because many susceptible agents pass through the infected agents compared to the minimum contact case.
We found that the final epidemic size became minimum when infected agents only move in one direction at the same speed as the other agents (Fig.~\ref{biasMovement}).

Figure \ref{sensitivityAnalysis}b shows the extinction time of disease in each setting.
It indicates the time that the disease completely dissapeared from the environment.
As also seen in Figs.\ref{noMovement} and \ref{highMovement}, the extinction time is basically longer when agents do not move.
When agents frequently move, the epidemic spreads to the entire population quickly and then those agents recover from the infection.
In contrast, when agents hardly move, the infection time is different for each agent.
Only when susceptive agents pass through infected agents do they get infected, resulting in the longer time of extinction.
This tendency is systematically confirmed in Fig.~\ref{sensitivityAnalysis}b.

\begin{figure}[!t]
  \centering
  \includegraphics[width=\columnwidth]{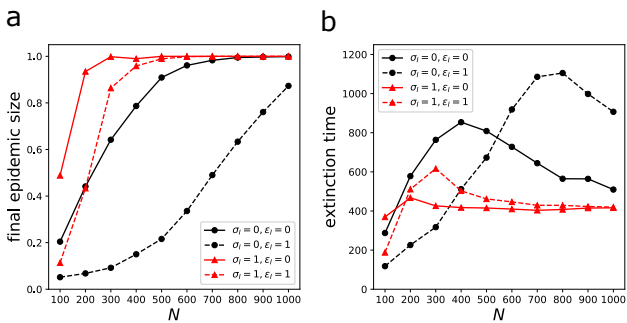}
  \caption{(Color online) The effect of  population size $N$ for the final epidemic size and the extinction time of disease.}
  \label{dependencyN}
 \end{figure}

Next, we varied the population size $N$ while the other parameters are fixed at $I_0=2, r=2$ and $\gamma=0.02$.
Figure \ref{dependencyN} shows the results.
As $N$ becomes larger, the final epidemic sizes increase.
Because the size of the space is fixed, the density becomes large as $N$ becomes larger.
For susceptible agents, high density leads to a higher chance of contact with infected agents.
This is why the final epidemic size is large when $N$ is large.
We focus on the extinction time when $N$ is varied (Fig.~\ref{dependencyN}b).
In the large population size ($N>600$), the extinction time became large when the final epidemic size was small, as seen in Fig.~\ref{sensitivityAnalysis}.
However, if the population is too small ($N \leq 300$), the minimum contact case ($\epsilon_I=0$ and $\sigma_I=1$) leads to the shortest extinction time.
In this setting, because the environment is sparse, contact between infected and susceptible rarely occurs.
Thus, further infection is frequently prevented in the minimum contact case, resulting in the shortest extinction time.
It should be noted that, when $N$ is varied, there are peaks in the middle for the extinction time.
Once the disease spreads to the entire population, it will disappear sooner or later.
In this case, the extinction time becomes relatively shorter.
It depends on each setting; $N \ge 400$ for $\sigma_I=1$, $N \ge 500$ for $\sigma_I=0$ and $\epsilon_I=0$, and $N \ge 900$ for $\sigma_I=0$ and $\epsilon_I=1$.

Finally, we focus on the effect of the infection radius $r$.
Obviously, as $r$ becomes larger, the final epidemic size increases.
The total tendency is the same with the other results.
The high movement case ($\sigma_I=1$ and $\epsilon_I=1$) raises the final epidemic size even when $r$ is relatively low.
When $\sigma_I=1$ and $\epsilon_I=0$, the final epidemic size raises even faster.
Conversely, the minimum contact case ($\sigma_I=0$ and $\epsilon_I=1$) slowly raises the final epidemic size.
For the extinction time, the peaks exist at the medium $r$.
These peaks correspond to the raised points for the final epidemic size.
Once the disease spreads to the entire population, agents recover from the disease quickly.
Thus, the peaks exist at the medium $r$.
 
   \begin{figure}[!t]
  \centering
  \includegraphics[width=\columnwidth]{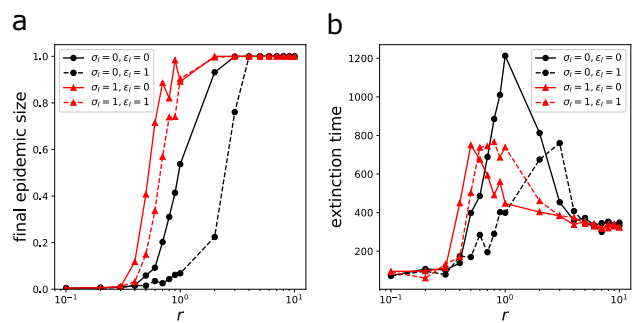}
  \caption{(Color online) The effect of  infection radius $r$ for the final epidemic size and the extinction time of disease.}
  \label{dependency_r}
 \end{figure}

 \section{Summary}
 \label{sec:Summary}
In this paper, we developed an agent-based model of susceptible-infected-recovered (SIR) epidemic process in a 2D continuous space.
In contrast to traditional epidemic models with random diffusion, we newly incorporated two features to the spatial epidemic system: biased movement and reduced mobility of infected agents.
We investigated the final epidemic size and extinction time when both movement bias and reduced mobility of infected individuals changed.
The results showed that when the movement of infected people is small, the final epidemic size becomes small.
However, that crucially depended on the movement type of agents.
Furthermore, the reduced mobility of infected agents lengthened the extinction time of the disease because the infection progressed slowly, even though the reduced mobility prevented the disease from spreading.
This fact suggests that isolating infected people is effective to prevent a pandemic.
However, if the duration of the isolation is not long enough, it may result in a long-lasting epidemic.
Therefore, governments need to take into account the nature of each disease (e.g. infection period, immunity, or infection distance) when they make policies.
 
 \section*{Acknowledgment}
 This work was supported from JSPS KAKENHI to G.I. (No. 17K17784).


\begin{thebibliography}{29}%
\makeatletter
\providecommand \@ifxundefined [1]{%
 \@ifx{#1\undefined}
}%
\providecommand \@ifnum [1]{%
 \ifnum #1\expandafter \@firstoftwo
 \else \expandafter \@secondoftwo
 \fi
}%
\providecommand \@ifx [1]{%
 \ifx #1\expandafter \@firstoftwo
 \else \expandafter \@secondoftwo
 \fi
}%
\providecommand \natexlab [1]{#1}%
\providecommand \enquote  [1]{``#1''}%
\providecommand \bibnamefont  [1]{#1}%
\providecommand \bibfnamefont [1]{#1}%
\providecommand \citenamefont [1]{#1}%
\providecommand \href@noop [0]{\@secondoftwo}%
\providecommand \href [0]{\begingroup \@sanitize@url \@href}%
\providecommand \@href[1]{\@@startlink{#1}\@@href}%
\providecommand \@@href[1]{\endgroup#1\@@endlink}%
\providecommand \@sanitize@url [0]{\catcode `\\12\catcode `\$12\catcode
  `\&12\catcode `\#12\catcode `\^12\catcode `\_12\catcode `\%12\relax}%
\providecommand \@@startlink[1]{}%
\providecommand \@@endlink[0]{}%
\providecommand \url  [0]{\begingroup\@sanitize@url \@url }%
\providecommand \@url [1]{\endgroup\@href {#1}{\urlprefix }}%
\providecommand \urlprefix  [0]{URL }%
\providecommand \Eprint [0]{\href }%
\providecommand \doibase [0]{http://dx.doi.org/}%
\providecommand \selectlanguage [0]{\@gobble}%
\providecommand \bibinfo  [0]{\@secondoftwo}%
\providecommand \bibfield  [0]{\@secondoftwo}%
\providecommand \translation [1]{[#1]}%
\providecommand \BibitemOpen [0]{}%
\providecommand \bibitemStop [0]{}%
\providecommand \bibitemNoStop [0]{.\EOS\space}%
\providecommand \EOS [0]{\spacefactor3000\relax}%
\providecommand \BibitemShut  [1]{\csname bibitem#1\endcsname}%
\let\auto@bib@innerbib\@empty
\bibitem [{\citenamefont {Kermac}\ and\ \citenamefont
  {McKendrick}(1927)}]{KermackMcKendrick_ProcRSocA_1927}%
  \BibitemOpen
  \bibfield  {author} {\bibinfo {author} {\bibfnamefont {W.~O.}\ \bibnamefont
  {Kermac}}\ and\ \bibinfo {author} {\bibfnamefont {A.~G.}\ \bibnamefont
  {McKendrick}},\ }\href {\doibase 10.1098/rspa.1927.0118} {\bibfield
  {journal} {\bibinfo  {journal} {Proc. R. Soc. A-Math. Phys. Eng. Sci.}\
  }\textbf {\bibinfo {volume} {115}},\ \bibinfo {pages} {700} (\bibinfo {year}
  {1927})}\BibitemShut {NoStop}%
\bibitem [{\citenamefont {Hethcote}(2000)}]{Hethcote_SIAMRev_2000}%
  \BibitemOpen
  \bibfield  {author} {\bibinfo {author} {\bibfnamefont {H.~W.}\ \bibnamefont
  {Hethcote}},\ }\href {\doibase 10.1137/S0036144500371907} {\bibfield
  {journal} {\bibinfo  {journal} {SIAM Rev.}\ }\textbf {\bibinfo {volume}
  {42}},\ \bibinfo {pages} {599} (\bibinfo {year} {2000})}\BibitemShut
  {NoStop}%
\bibitem [{\citenamefont {Diekmann}\ \emph {et~al.}(2012)\citenamefont
  {Diekmann}, \citenamefont {Heesterbeek},\ and\ \citenamefont
  {Britton}}]{Diekmann_etal2012}%
  \BibitemOpen
  \bibfield  {author} {\bibinfo {author} {\bibfnamefont {O.}~\bibnamefont
  {Diekmann}}, \bibinfo {author} {\bibfnamefont {H.}~\bibnamefont
  {Heesterbeek}}, \ and\ \bibinfo {author} {\bibfnamefont {T.}~\bibnamefont
  {Britton}},\ }\href@noop {} {\emph {\bibinfo {title} {Mathematical tools for
  understanding infectious disease dynamics}}}\ (\bibinfo  {publisher}
  {Princeton University Press, Princeton},\ \bibinfo {year} {2012})\BibitemShut
  {NoStop}%
\bibitem [{\citenamefont {Allen}\ \emph {et~al.}(2008)\citenamefont {Allen},
  \citenamefont {Bauch}, \citenamefont {Castillo-Chavez}, \citenamefont {Earn},
  \citenamefont {Feng},\ and\ \citenamefont {Lewis}}]{Allen_etal2008}%
  \BibitemOpen
  \bibfield  {author} {\bibinfo {author} {\bibfnamefont {L.}~\bibnamefont
  {Allen}}, \bibinfo {author} {\bibfnamefont {C.}~\bibnamefont {Bauch}},
  \bibinfo {author} {\bibfnamefont {C.}~\bibnamefont {Castillo-Chavez}},
  \bibinfo {author} {\bibfnamefont {D.}~\bibnamefont {Earn}}, \bibinfo {author}
  {\bibfnamefont {Z.}~\bibnamefont {Feng}}, \ and\ \bibinfo {author}
  {\bibfnamefont {M.}~\bibnamefont {Lewis}},\ }\href@noop {} {\emph {\bibinfo
  {title} {Mathematical epidemiology}}}\ (\bibinfo  {publisher} {Springer,
  Berlin},\ \bibinfo {year} {2008})\BibitemShut {NoStop}%
\bibitem [{\citenamefont {Stoddard}\ \emph {et~al.}(2009)\citenamefont
  {Stoddard}, \citenamefont {Morrison}, \citenamefont {Vazquez-Prokopec},
  \citenamefont {{Paz Soldan}}, \citenamefont {Kochel}, \citenamefont {Kitron},
  \citenamefont {Elder},\ and\ \citenamefont
  {Scott}}]{Steven_etal_PLoSNeglTropDis2009}%
  \BibitemOpen
  \bibfield  {author} {\bibinfo {author} {\bibfnamefont {S.~T.}\ \bibnamefont
  {Stoddard}}, \bibinfo {author} {\bibfnamefont {A.~C.}\ \bibnamefont
  {Morrison}}, \bibinfo {author} {\bibfnamefont {G.~M.}\ \bibnamefont
  {Vazquez-Prokopec}}, \bibinfo {author} {\bibfnamefont {V.}~\bibnamefont {{Paz
  Soldan}}}, \bibinfo {author} {\bibfnamefont {T.~J.}\ \bibnamefont {Kochel}},
  \bibinfo {author} {\bibfnamefont {U.}~\bibnamefont {Kitron}}, \bibinfo
  {author} {\bibfnamefont {J.~P.}\ \bibnamefont {Elder}}, \ and\ \bibinfo
  {author} {\bibfnamefont {T.~W.}\ \bibnamefont {Scott}},\ }\href {\doibase
  10.1371/journal.pntd.0000481} {\bibfield  {journal} {\bibinfo  {journal}
  {PLoS Negl. Trop. Dis.}\ }\textbf {\bibinfo {volume} {3}},\ \bibinfo {pages}
  {e481} (\bibinfo {year} {2009})}\BibitemShut {NoStop}%
\bibitem [{\citenamefont {Barmak}\ \emph {et~al.}(2011)\citenamefont {Barmak},
  \citenamefont {Dorso}, \citenamefont {Otero},\ and\ \citenamefont
  {Solari}}]{Barmak_etal_PhysRevE2011}%
  \BibitemOpen
  \bibfield  {author} {\bibinfo {author} {\bibfnamefont {D.~H.}\ \bibnamefont
  {Barmak}}, \bibinfo {author} {\bibfnamefont {C.~O.}\ \bibnamefont {Dorso}},
  \bibinfo {author} {\bibfnamefont {M.}~\bibnamefont {Otero}}, \ and\ \bibinfo
  {author} {\bibfnamefont {H.~G.}\ \bibnamefont {Solari}},\ }\href {\doibase
  10.1103/PhysRevE.84.011901} {\bibfield  {journal} {\bibinfo  {journal} {Phys.
  Rev. E}\ }\textbf {\bibinfo {volume} {84}},\ \bibinfo {pages} {011901}
  (\bibinfo {year} {2011})}\BibitemShut {NoStop}%
\bibitem [{\citenamefont {Li}\ \emph {et~al.}(2010)\citenamefont {Li},
  \citenamefont {Cao},\ and\ \citenamefont {Cao}}]{Li_etal_EPJB2010}%
  \BibitemOpen
  \bibfield  {author} {\bibinfo {author} {\bibfnamefont {X.}~\bibnamefont
  {Li}}, \bibinfo {author} {\bibfnamefont {L.}~\bibnamefont {Cao}}, \ and\
  \bibinfo {author} {\bibfnamefont {G.~F.}\ \bibnamefont {Cao}},\ }\href
  {\doibase 10.1140/epjb/e2010-00090-9} {\bibfield  {journal} {\bibinfo
  {journal} {Eur. Phys. J. B}\ }\textbf {\bibinfo {volume} {75}},\ \bibinfo
  {pages} {319} (\bibinfo {year} {2010})}\BibitemShut {NoStop}%
\bibitem [{\citenamefont {Zhou}\ \emph {et~al.}(2012)\citenamefont {Zhou},
  \citenamefont {Chung}, \citenamefont {Chew},\ and\ \citenamefont
  {Lai}}]{Zhou_etal_PhysRevE2012}%
  \BibitemOpen
  \bibfield  {author} {\bibinfo {author} {\bibfnamefont {J.}~\bibnamefont
  {Zhou}}, \bibinfo {author} {\bibfnamefont {N.~N.}\ \bibnamefont {Chung}},
  \bibinfo {author} {\bibfnamefont {L.~Y.}\ \bibnamefont {Chew}}, \ and\
  \bibinfo {author} {\bibfnamefont {C.~H.}\ \bibnamefont {Lai}},\ }\href
  {\doibase 10.1103/PhysRevE.86.026115} {\bibfield  {journal} {\bibinfo
  {journal} {Phys. Rev. E}\ }\textbf {\bibinfo {volume} {86}},\ \bibinfo
  {pages} {026115} (\bibinfo {year} {2012})}\BibitemShut {NoStop}%
\bibitem [{\citenamefont {Hong}\ \emph {et~al.}(2016)\citenamefont {Hong},
  \citenamefont {Yang}, \citenamefont {Zhao},\ and\ \citenamefont
  {Ma}}]{Hong_etal_IntJSystSci2016}%
  \BibitemOpen
  \bibfield  {author} {\bibinfo {author} {\bibfnamefont {S.}~\bibnamefont
  {Hong}}, \bibinfo {author} {\bibfnamefont {H.}~\bibnamefont {Yang}}, \bibinfo
  {author} {\bibfnamefont {T.}~\bibnamefont {Zhao}}, \ and\ \bibinfo {author}
  {\bibfnamefont {X.}~\bibnamefont {Ma}},\ }\href {\doibase
  10.1080/00207721.2015.1022890} {\bibfield  {journal} {\bibinfo  {journal}
  {Int. J. Syst. Sci.}\ }\textbf {\bibinfo {volume} {47}},\ \bibinfo {pages}
  {2745} (\bibinfo {year} {2016})}\BibitemShut {NoStop}%
\bibitem [{\citenamefont {Buscarino}\ \emph {et~al.}(2014)\citenamefont
  {Buscarino}, \citenamefont {Fortuna}, \citenamefont {Frasca},\ and\
  \citenamefont {Rizzo}}]{Buscarino_etal_PhysRevE2014}%
  \BibitemOpen
  \bibfield  {author} {\bibinfo {author} {\bibfnamefont {A.}~\bibnamefont
  {Buscarino}}, \bibinfo {author} {\bibfnamefont {L.}~\bibnamefont {Fortuna}},
  \bibinfo {author} {\bibfnamefont {M.}~\bibnamefont {Frasca}}, \ and\ \bibinfo
  {author} {\bibfnamefont {A.}~\bibnamefont {Rizzo}},\ }\href {\doibase
  10.1103/PhysRevE.90.042813} {\bibfield  {journal} {\bibinfo  {journal} {Phys.
  Rev. E}\ }\textbf {\bibinfo {volume} {90}},\ \bibinfo {pages} {042813}
  (\bibinfo {year} {2014})}\BibitemShut {NoStop}%
\bibitem [{\citenamefont {Gong}\ \emph {et~al.}(2014)\citenamefont {Gong},
  \citenamefont {Song},\ and\ \citenamefont {Jiang}}]{Gong_etal_PhysA2014}%
  \BibitemOpen
  \bibfield  {author} {\bibinfo {author} {\bibfnamefont {Y.-W.}\ \bibnamefont
  {Gong}}, \bibinfo {author} {\bibfnamefont {Y.-R.}\ \bibnamefont {Song}}, \
  and\ \bibinfo {author} {\bibfnamefont {G.-P.}\ \bibnamefont {Jiang}},\ }\href
  {\doibase 10.1016/j.physa.2014.08.056} {\bibfield  {journal} {\bibinfo
  {journal} {Physica A}\ }\textbf {\bibinfo {volume} {416}},\ \bibinfo {pages}
  {208} (\bibinfo {year} {2014})}\BibitemShut {NoStop}%
\bibitem [{\citenamefont {Barmak}\ \emph {et~al.}(2016)\citenamefont {Barmak},
  \citenamefont {Dorso},\ and\ \citenamefont {Otero}}]{Barmak_etal_PhysA2016}%
  \BibitemOpen
  \bibfield  {author} {\bibinfo {author} {\bibfnamefont {D.~H.}\ \bibnamefont
  {Barmak}}, \bibinfo {author} {\bibfnamefont {C.~O.}\ \bibnamefont {Dorso}}, \
  and\ \bibinfo {author} {\bibfnamefont {M.}~\bibnamefont {Otero}},\ }\href
  {\doibase 10.1016/j.physa.2015.12.015} {\bibfield  {journal} {\bibinfo
  {journal} {Physica A}\ }\textbf {\bibinfo {volume} {447}},\ \bibinfo {pages}
  {129} (\bibinfo {year} {2016})}\BibitemShut {NoStop}%
\bibitem [{\citenamefont {Li}\ \emph {et~al.}(2015{\natexlab{a}})\citenamefont
  {Li}, \citenamefont {Yu}, \citenamefont {Zeng}, \citenamefont {Ding},\ and\
  \citenamefont {Ma}}]{Li_etal_CommunNonlinearSci2015}%
  \BibitemOpen
  \bibfield  {author} {\bibinfo {author} {\bibfnamefont {K.}~\bibnamefont
  {Li}}, \bibinfo {author} {\bibfnamefont {H.}~\bibnamefont {Yu}}, \bibinfo
  {author} {\bibfnamefont {Z.}~\bibnamefont {Zeng}}, \bibinfo {author}
  {\bibfnamefont {Y.}~\bibnamefont {Ding}}, \ and\ \bibinfo {author}
  {\bibfnamefont {Z.}~\bibnamefont {Ma}},\ }\href {\doibase
  10.1016/j.cnsns.2014.07.016} {\bibfield  {journal} {\bibinfo  {journal}
  {Commun. Nonlinear. Sci. Numer. Simul.}\ }\textbf {\bibinfo {volume} {22}},\
  \bibinfo {pages} {596} (\bibinfo {year} {2015}{\natexlab{a}})}\BibitemShut
  {NoStop}%
\bibitem [{\citenamefont {Buscarino}\ \emph {et~al.}(2010)\citenamefont
  {Buscarino}, \citenamefont {Stefano}, \citenamefont {Fortuna}, \citenamefont
  {Frasca},\ and\ \citenamefont
  {Latora}}]{Buscarino_etal_IntJBifurcationChaos2010}%
  \BibitemOpen
  \bibfield  {author} {\bibinfo {author} {\bibfnamefont {A.}~\bibnamefont
  {Buscarino}}, \bibinfo {author} {\bibfnamefont {A.~D.}\ \bibnamefont
  {Stefano}}, \bibinfo {author} {\bibfnamefont {L.}~\bibnamefont {Fortuna}},
  \bibinfo {author} {\bibfnamefont {M.}~\bibnamefont {Frasca}}, \ and\ \bibinfo
  {author} {\bibfnamefont {V.}~\bibnamefont {Latora}},\ }\href {\doibase
  10.1142/S0218127410026058} {\bibfield  {journal} {\bibinfo  {journal} {Int.
  J. Bifurcation Chaos}\ }\textbf {\bibinfo {volume} {20}},\ \bibinfo {pages}
  {765} (\bibinfo {year} {2010})}\BibitemShut {NoStop}%
\bibitem [{\citenamefont {Nagatani}\ \emph {et~al.}(2018)\citenamefont
  {Nagatani}, \citenamefont {Ichinose},\ and\ \citenamefont
  {Tainaka}}]{Nagatani_etal_JTheorBiol2018}%
  \BibitemOpen
  \bibfield  {author} {\bibinfo {author} {\bibfnamefont {T.}~\bibnamefont
  {Nagatani}}, \bibinfo {author} {\bibfnamefont {G.}~\bibnamefont {Ichinose}},
  \ and\ \bibinfo {author} {\bibfnamefont {K.}~\bibnamefont {Tainaka}},\ }\href
  {\doibase 10.1016/j.jtbi.2018.04.029} {\bibfield  {journal} {\bibinfo
  {journal} {J. Theor. Biol.}\ }\textbf {\bibinfo {volume} {450}},\ \bibinfo
  {pages} {66} (\bibinfo {year} {2018})}\BibitemShut {NoStop}%
\bibitem [{\citenamefont {Huang}\ \emph
  {et~al.}(2016{\natexlab{a}})\citenamefont {Huang}, \citenamefont {Ding},
  \citenamefont {Feng},\ and\ \citenamefont {Pan}}]{Huang_etal_JStatMechl2016}%
  \BibitemOpen
  \bibfield  {author} {\bibinfo {author} {\bibfnamefont {Y.}~\bibnamefont
  {Huang}}, \bibinfo {author} {\bibfnamefont {L.}~\bibnamefont {Ding}},
  \bibinfo {author} {\bibfnamefont {Y.}~\bibnamefont {Feng}}, \ and\ \bibinfo
  {author} {\bibfnamefont {J.}~\bibnamefont {Pan}},\ }\href {\doibase
  10.1088/1742-5468/2016/10/103501} {\bibfield  {journal} {\bibinfo  {journal}
  {J. Stat. Mech}\ }\textbf {\bibinfo {volume} {2016}},\ \bibinfo {pages}
  {103501} (\bibinfo {year} {2016}{\natexlab{a}})}\BibitemShut {NoStop}%
\bibitem [{\citenamefont {Sharma}\ and\ \citenamefont
  {Gupta}(2017)}]{SharmaGupta_PhysAl2017}%
  \BibitemOpen
  \bibfield  {author} {\bibinfo {author} {\bibfnamefont {N.}~\bibnamefont
  {Sharma}}\ and\ \bibinfo {author} {\bibfnamefont {A.~K.}\ \bibnamefont
  {Gupta}},\ }\href {\doibase 10.1016/j.physa.2016.12.010} {\bibfield
  {journal} {\bibinfo  {journal} {Physica A}\ }\textbf {\bibinfo {volume}
  {471}},\ \bibinfo {pages} {114} (\bibinfo {year} {2017})}\BibitemShut
  {NoStop}%
\bibitem [{\citenamefont {Hosono}\ and\ \citenamefont
  {Ilyas}(1995)}]{HosonoIlyas_MathModelsMethods1995}%
  \BibitemOpen
  \bibfield  {author} {\bibinfo {author} {\bibfnamefont {Y.}~\bibnamefont
  {Hosono}}\ and\ \bibinfo {author} {\bibfnamefont {B.}~\bibnamefont {Ilyas}},\
  }\href {\doibase 10.1142/S0218202595000504} {\bibfield  {journal} {\bibinfo
  {journal} {Math. Models Methods Appl. Sci.}\ }\textbf {\bibinfo {volume}
  {5}},\ \bibinfo {pages} {935} (\bibinfo {year} {1995})}\BibitemShut {NoStop}%
\bibitem [{\citenamefont {Wang}\ \emph {et~al.}(2012)\citenamefont {Wang},
  \citenamefont {Wang},\ and\ \citenamefont {Wu}}]{Wang_etal_DisConDynSys2012}%
  \BibitemOpen
  \bibfield  {author} {\bibinfo {author} {\bibfnamefont {X.~S.}\ \bibnamefont
  {Wang}}, \bibinfo {author} {\bibfnamefont {H.}~\bibnamefont {Wang}}, \ and\
  \bibinfo {author} {\bibfnamefont {J.}~\bibnamefont {Wu}},\ }\href {\doibase
  10.3934/dcds.2012.32.3303} {\bibfield  {journal} {\bibinfo  {journal}
  {Discrete Continuous Dyn Syst Ser A}\ }\textbf {\bibinfo {volume} {32}},\
  \bibinfo {pages} {3303} (\bibinfo {year} {2012})}\BibitemShut {NoStop}%
\bibitem [{\citenamefont {Li}\ \emph {et~al.}(2015{\natexlab{b}})\citenamefont
  {Li}, \citenamefont {Li},\ and\ \citenamefont
  {Lin}}]{Li_etal_CommPureApplAnal2015}%
  \BibitemOpen
  \bibfield  {author} {\bibinfo {author} {\bibfnamefont {Y.}~\bibnamefont
  {Li}}, \bibinfo {author} {\bibfnamefont {W.-T.}\ \bibnamefont {Li}}, \ and\
  \bibinfo {author} {\bibfnamefont {G.}~\bibnamefont {Lin}},\ }\href {\doibase
  10.3934/cpaa.2015.14.1001} {\bibfield  {journal} {\bibinfo  {journal}
  {Commun. Pur. Appl. Anal.}\ }\textbf {\bibinfo {volume} {14}},\ \bibinfo
  {pages} {1001} (\bibinfo {year} {2015}{\natexlab{b}})}\BibitemShut {NoStop}%
\bibitem [{\citenamefont {Bai}\ and\ \citenamefont
  {Zhang}(2015)}]{BaiZhang_CommNonlinSciNumerSimul2015}%
  \BibitemOpen
  \bibfield  {author} {\bibinfo {author} {\bibfnamefont {Z.}~\bibnamefont
  {Bai}}\ and\ \bibinfo {author} {\bibfnamefont {S.}~\bibnamefont {Zhang}},\
  }\href {\doibase https://doi.org/10.1016/j.cnsns.2014.07.005} {\bibfield
  {journal} {\bibinfo  {journal} {Commun. Nonlinear. Sci. Numer. Simul.}\
  }\textbf {\bibinfo {volume} {22}},\ \bibinfo {pages} {1370} (\bibinfo {year}
  {2015})}\BibitemShut {NoStop}%
\bibitem [{\citenamefont {Beardmore}\ and\ \citenamefont
  {Beardmore}(2003)}]{BeardmoreBeardmore_ProcRSocLondA2003}%
  \BibitemOpen
  \bibfield  {author} {\bibinfo {author} {\bibfnamefont {I.}~\bibnamefont
  {Beardmore}}\ and\ \bibinfo {author} {\bibfnamefont {R.}~\bibnamefont
  {Beardmore}},\ }\href {\doibase 10.1098/rspa.2002.1080} {\bibfield  {journal}
  {\bibinfo  {journal} {Proc. Math. Phys. Eng. Sci.}\ }\textbf {\bibinfo
  {volume} {459}},\ \bibinfo {pages} {1427} (\bibinfo {year}
  {2003})}\BibitemShut {NoStop}%
\bibitem [{\citenamefont {Gudelj}\ and\ \citenamefont
  {White}(2004)}]{GudeljWhite_TheorPopulBiol2004}%
  \BibitemOpen
  \bibfield  {author} {\bibinfo {author} {\bibfnamefont {I.}~\bibnamefont
  {Gudelj}}\ and\ \bibinfo {author} {\bibfnamefont {K.~A.~J.}\ \bibnamefont
  {White}},\ }\href {\doibase https://doi.org/10.1016/j.tpb.2004.04.003}
  {\bibfield  {journal} {\bibinfo  {journal} {Theor. Popul. Biol.}\ }\textbf
  {\bibinfo {volume} {66}},\ \bibinfo {pages} {139} (\bibinfo {year}
  {2004})}\BibitemShut {NoStop}%
\bibitem [{\citenamefont {Gudelj}\ \emph {et~al.}(2004)\citenamefont {Gudelj},
  \citenamefont {White},\ and\ \citenamefont
  {Britton}}]{Gudelj_etal_BullMathBiol2004}%
  \BibitemOpen
  \bibfield  {author} {\bibinfo {author} {\bibfnamefont {I.}~\bibnamefont
  {Gudelj}}, \bibinfo {author} {\bibfnamefont {K.~A.~J.}\ \bibnamefont
  {White}}, \ and\ \bibinfo {author} {\bibfnamefont {N.~F.}\ \bibnamefont
  {Britton}},\ }\href {\doibase 10.1016/S0092-8240(03)00075-2} {\bibfield
  {journal} {\bibinfo  {journal} {Bull. Math. Biol.}\ }\textbf {\bibinfo
  {volume} {66}},\ \bibinfo {pages} {91} (\bibinfo {year} {2004})}\BibitemShut
  {NoStop}%
\bibitem [{\citenamefont {Zhou}\ and\ \citenamefont
  {Liu}(2009{\natexlab{a}})}]{ZhouLiu_PhysA2009}%
  \BibitemOpen
  \bibfield  {author} {\bibinfo {author} {\bibfnamefont {J.}~\bibnamefont
  {Zhou}}\ and\ \bibinfo {author} {\bibfnamefont {Z.}~\bibnamefont {Liu}},\
  }\href {\doibase https://doi.org/10.1016/j.physa.2008.12.014} {\bibfield
  {journal} {\bibinfo  {journal} {Physica A}\ }\textbf {\bibinfo {volume}
  {388}},\ \bibinfo {pages} {1228} (\bibinfo {year}
  {2009}{\natexlab{a}})}\BibitemShut {NoStop}%
\bibitem [{\citenamefont {Huang}\ \emph
  {et~al.}(2016{\natexlab{b}})\citenamefont {Huang}, \citenamefont {Ding},
  \citenamefont {Feng},\ and\ \citenamefont {Pan}}]{Pan_JSTAT2016}%
  \BibitemOpen
  \bibfield  {author} {\bibinfo {author} {\bibfnamefont {Y.}~\bibnamefont
  {Huang}}, \bibinfo {author} {\bibfnamefont {L.}~\bibnamefont {Ding}},
  \bibinfo {author} {\bibfnamefont {Y.}~\bibnamefont {Feng}}, \ and\ \bibinfo
  {author} {\bibfnamefont {J.}~\bibnamefont {Pan}},\ }\href@noop {} {\bibfield
  {journal} {\bibinfo  {journal} {J. Stat. Mech}\ }\textbf {\bibinfo {volume}
  {2016}},\ \bibinfo {pages} {103501} (\bibinfo {year}
  {2016}{\natexlab{b}})}\BibitemShut {NoStop}%
\bibitem [{\citenamefont {Zhou}\ and\ \citenamefont
  {Liu}(2009{\natexlab{b}})}]{Zhou_PhysA2009}%
  \BibitemOpen
  \bibfield  {author} {\bibinfo {author} {\bibfnamefont {J.}~\bibnamefont
  {Zhou}}\ and\ \bibinfo {author} {\bibfnamefont {Z.}~\bibnamefont {Liu}},\
  }\href {\doibase https://doi.org/10.1016/j.physa.2008.12.014} {\bibfield
  {journal} {\bibinfo  {journal} {Physica A}\ }\textbf {\bibinfo {volume}
  {388}},\ \bibinfo {pages} {1228} (\bibinfo {year}
  {2009}{\natexlab{b}})}\BibitemShut {NoStop}%
\bibitem [{\citenamefont {Buscarino}\ \emph {et~al.}(2008)\citenamefont
  {Buscarino}, \citenamefont {Fortuna}, \citenamefont {Frasca},\ and\
  \citenamefont {Latora}}]{Buscarino_EPL2008}%
  \BibitemOpen
  \bibfield  {author} {\bibinfo {author} {\bibfnamefont {A.}~\bibnamefont
  {Buscarino}}, \bibinfo {author} {\bibfnamefont {L.}~\bibnamefont {Fortuna}},
  \bibinfo {author} {\bibfnamefont {M.}~\bibnamefont {Frasca}}, \ and\ \bibinfo
  {author} {\bibfnamefont {V.}~\bibnamefont {Latora}},\ }\href@noop {}
  {\bibfield  {journal} {\bibinfo  {journal} {EPL}\ }\textbf {\bibinfo {volume}
  {82}},\ \bibinfo {pages} {38002} (\bibinfo {year} {2008})}\BibitemShut
  {NoStop}%
\bibitem [{\citenamefont {Nagatani}\ \emph {et~al.}(2017)\citenamefont
  {Nagatani}, \citenamefont {Ichinose},\ and\ \citenamefont
  {Tainaka}}]{Nagatani_etal_JPSJ2017}%
  \BibitemOpen
  \bibfield  {author} {\bibinfo {author} {\bibfnamefont {T.}~\bibnamefont
  {Nagatani}}, \bibinfo {author} {\bibfnamefont {G.}~\bibnamefont {Ichinose}},
  \ and\ \bibinfo {author} {\bibfnamefont {K.}~\bibnamefont {Tainaka}},\ }\href
  {\doibase 10.7566/JPSJ.86.113001} {\bibfield  {journal} {\bibinfo  {journal}
  {Journal of the Physical Society of Japan}\ }\textbf {\bibinfo {volume}
  {86}},\ \bibinfo {pages} {113001} (\bibinfo {year} {2017})}\BibitemShut
  {NoStop}%
\end{thebibliography}
\providecommand{\noopsort}[1]{}\providecommand{\singleletter}[1]{#1}%

\end{document}